\documentclass[letterpaper]{article} 
\usepackage{aaai2026}  
\nocopyright
\usepackage{times}  
\usepackage{helvet}  
\usepackage{courier}  
\usepackage[hyphens]{url}  
\usepackage{graphicx} 
\usepackage{natbib}  
\usepackage{caption} 
\usepackage{amsmath}
\usepackage{amsfonts}
\usepackage{subcaption}
\usepackage{booktabs}

\DeclareCaptionStyle{ruled}{labelfont=normalfont,labelsep=colon,strut=off} 

\title{Taking the Whys Seriously: Limitations of Counterfactual Explanations in Justification and Recourse}
\author {
    Mattia Cerrato\textsuperscript{\rm 1}\equalcontrib,
    Otto Sahlgren\textsuperscript{\rm 2}\equalcontrib,
    Xenia Heilmann\textsuperscript{\rm 1}
}
\affiliations {
    \textsuperscript{\rm 1}Johannes Gutenberg University Mainz, Germany\\
    \textsuperscript{\rm 2}University of Tampere, Finland\\
    mcerrato@uni-mainz.de, otto.sahlgren@tuni.fi, xheilmann@uni-mainz.de
}

\begin{document}

\maketitle

\begin{abstract}
Counterfactual explanations (CEs) are widely used in explainable artificial intelligence (AI) to show how a model's outputs would change if the input features were manipulated. This technique is used for a range of tasks such as debugging models, explaining predictions, justifying decisions, and providing algorithmic recourse. In this paper, we explore the normative legitimacy of employing counterfactuals in real-life model deployment settings. We discuss the different stakes involved in these different purposes for which CEs are commonly employed, and find stricter requirements for justification and recourse. In particular, we find that naive application of CEs for justification and recourse can lead to ignoring contestable choices made throughout the machine learning (ML) pipeline, thus obfuscating that decisions and counterfactuals for those decisions are also artifacts of an organization's materialized design and governance choices. 
We demonstrate this with four empirical experiments involving interventions at stages of the ML pipeline ``upstream" of the explanation itself, and show that these affect the generated counterfactuals. We find that an organization's choices on measurement models for feature and labels, business requirements, model validation, and the metric of model success have as much or more impact on the generated counterfactuals as the specifics of the generating method. Our findings underline the need to account for such choices upon providing justification and recourse, providing a stark reminder of the relational nature of these tasks. As putative justifications or recourse recommendations, CEs do not provide adequate answers to some important "why"-questions because they preclude consideration of whether the decision-maker ought to have acted differently.  
\end{abstract}

\section{Introduction}






The growing adoption of machine learning (ML) systems in high-stakes decision-making contexts has spurred research on the interrelated notions of transparency, explainability, and interpretability. Transparency, generally defined as “the availability of information about an actor that allows other actors to monitor the workings or performance of the first actor” \cite[p.430]{meijer2013transparency}, lacks a unique and well-defined equivalent in the context of ML systems. Instead, multiple notions of “algorithmic transparency” can be defined in relation to different objects and subcomponents as subject to transparency, where the transparency thereof can take different forms and gradations \cite[p.199]{diakopoulos2025prospective}. 
For example, “algorithmic transparency” can refer to the code, datasets, algorithms or their purposes and use-cases being disclosed in an understandable form to some audience.

One of the most celebrated techniques to this end is the generation of counterfactual explanations (CEs) \cite{wachter2017counterfactual,kusner2017counterfactual}, which have been referred to as ``de-facto standard in explainable artificial intelligence'' \cite{sokol2019desiderata}. 
The basic concept of CEs is to provide explanation to the ``why'' question: \emph{why does the model return this decision in this situation}?
CEs propose modifications to the input data that would incur a different decision by the model \cite{verma2024counterfactual}. 
Thus, CEs relate to \emph{singular decisions} rather than \emph{global approximation} of a complex model, and may provide an explanation by contrast rather than by local approximation as done by other techniques such as local explanations \cite{ribeiro2016should}.
CE is both attractive and flexible, seeing usage beyond the abstract opening of the ``black box'', such as in model debugging \cite{abid2022meaningfully}, and the provision of justification and recourse for algorithmic decisions \cite{karimi2021algorithmic,wachter2017counterfactual}.

Our contribution is motivated by critical research on CE and its limits in the aforementioned tasks. 
Though we acknowledge the potential of CEs, we are interested in gauging the extent to which they can reliably support justification, contestation, and reversal of decisions across real-world decision-making settings. 
Specifically, we believe that assessing CEs' sensitivity to expectable circumstantial variation is crucial for determining whether CE can \textit{robustly} realize values central to practices of public justification and recourse mechanisms, such as agency, trust, and accountability. For example, adequate recourse mechanisms will enable contestation, appeal, and reversal across a range of circumstances rather than only under narrowly specified conditions \cite{venkatasubramanian2020philosophical}. It is therefore crucial to assess whether CEs can deliver stable and meaningful recourse under variation characteristic to real-world decision-making settings.

Our paper provides both empirical and theoretical contributions. While prior research has explored the robustness of CEs under minor changes in the data, model, and generation method \cite{virgolin2022on,slack2021counterfactual,brughmans2024nice}, we measure the sensitivity of CEs under systematic ``upstream interventions" affecting the pipeline that starts from data collection, goes through model selection and fitting, and finally ends in the generation of CEs. Specifically, we start from explicitly modeling certain choices usually made on the ``business side'' of the ML pipeline as interventions on a structural causal model (SCM) representing the pipeline itself, such as the observation model of features, the labeling mechanism and the metric employed to operationalize the fitness of a model. For simplicity, we refer to these choices collectively as \textit{upstream choices} which produce different specifications of the SCM and, by extension, models. 

We find that upstream interventions can dramatically affect the CEs generated downstream, producing differences comparable in magnitude to those observed when the explanation method itself is changed \citep{brughmans2024nice}. Drawing on these findings, we substantiate the theoretical claim that several assumptions need to hold---in both an empirical and normative sense---for CEs to represent adequate ``whys'' in many real-life cases. Upstream interventions represent meaningful, non-neutral, and potentially contestable choices made by model providers or decision-makers. However, they remain unaccounted for when CEs are presented as \textit{evidence} to justify models or particular outputs or as \textit{recommendations} for decision reversal. We argue that taking the ``whys" seriously requires, besides mere explainability, broader transparency about upstream assumptions and choices, accordingly, and recognizing model providers and deployers as parties whose choices and policies require justification and potential revision.

\section{Background and Related Work}

 The concepts explainability and interpretability, often used interchangeably, have received significant attention in the ML literature \cite{lipton2018mythos}. Broadly, explainability refers to the ``the ability to explain or to present [the algorithm’s prediction logic] in understandable terms to a human'' \cite[p.2]{doshivelez2017towards}, whereas interpretability refers more specifically to “the degree to which an observer can understand the cause of a decision” \cite[p.8]{miller2019explanation} (see also \citet{biran2017explanation}). Technical approaches in these areas, which we describe below, pursue a range of overlapping but conceptually distinct aims. Examples include debugging, evaluating, and monitoring ML system performance; enabling decision-makers to explain and justify decisions to decision-subjects; ensuring that subjects and oversight bodies can understand, contest, trust those decisions; and providing subjects with actionable means of recourse in case of incorrect or unfavorable decisions \cite{diakopoulos2025prospective, verma2024counterfactual}. These concerns are reflected in existing regulations and literature on the so-called right to explanation \cite{juliussen2025right}. Article 22 of the General Data Protection Regulation (GDPR) states that “[t]he controller shall provide the data subject with [...] meaningful information about the logic involved, as well as the significance and the envisaged consequences of such processing for the data subject" \cite[Art. 21]{gdpr}. In addition, article 86 of the AI act states that, with some exceptions, “[a]ny affected person subject to a decision which is taken by the deployer on the basis of the output from a high-risk AI system” as classified in the AI act \cite[Art. 86]{AIA2024} “shall have the right to obtain from the deployer clear and meaningful explanations of the role of the AI system in the decision-making procedure and the main elements of the decision taken”.

Technical solutions to these ends may be grouped roughly into two categories. First, \textit{global explanation methods} seek to approximate a complex model in its entirety by identifying important input variables, for example, or by providing visual summaries \cite{lundberg2017unified}. Examples of global techniques include ones for computing feature importance and dependency plots. Second, \textit{local explanation methods} aim to explain individual decisions, either by approximating the model’s decision surface (i.e., a boundary or the regression hyperplane) in a region close to the decision point or by providing an alternative scenario where the model’s decision would be different. Local approximation tools such as LIME fall into the former class \cite{ribeiro2016should}, while CEs, which comprise our focus, fall under the latter. A further alternative worth mentioning is to employ statistical models which are transparent “off-the-shelf” whenever possible, as proposed by \citet{rudin2019stop}. Nonetheless, so-called “post-hoc” explainability has remained a central topic in AI research for over a decade \cite{guidotti2018survey}.

\subsection{Counterfactual Explanations (CE)}

This paper focuses on counterfactual explanation (CE), a class of local explanations that describe how a prediction model’s prediction (e.g., a classification or a ranking) would change given a change in the input \cite{verma2024counterfactual}. In plain terms, CE answers ``why" questions by computing “what ifs”, namely, by identifying and specifying minimal changes in input that would flip the output. How much job experience should a rejected job applicant have had in order to get hired? Would a change in an applicant’s salary have made a difference to their credit score? CEs are versatile and relatively easy to implement as they do not require opening ``black box" models \cite{wachter2017counterfactual}. They have become the “de facto standard in explainable artificial intelligence” \cite[p.2]{sokol2019desiderata}, accordingly. Subsequent sections discuss its numerous use-applications ranging from model debugging \cite{abid2022meaningfully} and model evaluation and justification (e.g., \citet{goethals2023precof,wang2024counterfactual}; see also \citet{kusner2017counterfactual}) to the provision of recourse \cite{wachter2017counterfactual,verma2024counterfactual}.

Formally, a counterfactual explanation $x'$ is an alternative input instance $x, x' \in \mathcal{X}$ which induces a different decision from a decision model $F$. Usually, the counterfactual instance $x'$ should be taken to be as similar as possible to $x$; sometimes, it is constrained to be an actual instance from the collected dataset $\mathcal{D} = \{ x_i, y_i \}^n$. Finding counterfactuals can then be operationalized as an optimization problem, of which various variants exist. For expository purposes, we consider the following problem: 

\begin{equation}
    \arg min_{X'} \mathcal{L}(F(X), F(X')) + \lambda \cdot d(X, X')
\end{equation}

where $\mathcal{L}$ is a loss (or cost) function modelling the similarity between $F(x)$ and $F(x')$, which should be minimized under the constraint that $F(x) \neq F(x')$; and $d(x, x')$ is a dissimilarity function computed directly on the instances (for instance, a norm defined in $\mathcal{X}^m$ where $m$ is the number of features available for each data point). Finally, $\lambda$ is a parameter available to the method user to balance out the two objectives.  

Methodologies for obtaining CEs differ in their underlying optimization techniques and their concern.  Advancement in CE research and evaluation of methodologies may be performed over different dimensions and under various concerns. We summarize here four key concerns central to our present discussion. We refer readers interested in an exhaustive list of evaluation criteria to the surveys by \citet{verma2024counterfactual} and \citet{bayrak2024evaluation}: 

\begin{itemize}
    \item \textbf{Validity:} a counterfactual explanation is considered valid if the counterfactual $x'$ is predicted by the model $f$ in a way considered desirable for counterfactual reasoning. As an example, if a user is wondering why their resumé was not selected after screening – that is, say, $y = f(x) = 0$, in the counterfactual explanation, $y' = f(x')$ should be equal to 1. That is, when considering $f$ as a classifier, counterfactual explanations need to cross the decision boundary. 

    \item \textbf{Minimality:} a minimal counterfactual explanation $x'$ is as close as possible to the explanandum $x$. This property is sometimes referred to as sparsity, which implies the minimization of the L1-norm $\left| x' - x \right|_1$ and mandates additionally that the least number of entries are changed. An appropriate norm choice is usually dependent on the feature types available. 

    \item \textbf{Actionability:} a counterfactual explanation is considered actionable if it does not include immutable, sensitive or otherwise hard-to-modify characteristics of the input data. Operationally, this may be implemented by a set of no-go features $\left|\mathcal{A}\right| \leq m$ which should not be included in the counterfactual explanation. The choice of $A$ may also be related to fairness concerns, broadly intended. As an example, under the constraint of actionability, an explanation in the resume’ screening classification should not include their gender or age. Under a cost-sensitive relaxation of actionability, it may be considered more actionable to suggest gaining some professional certification relevant to the job position rather than enrolling in a multi-year education program. 

    \item \textbf{Robustness:} a counterfactual explanation is considered robust if it remains valid under small perturbations to the model or its underlying data. \citet{fokkema2022noise} show that naive counterfactual methods applied to ensemble models can have validity below 50\% after model retraining, and that requiring higher prediction confidence for the counterfactual class mitigates this issue at the cost of increased distance from the original instance.
\end{itemize}

\subsection{Algorithmic Recourse (AR)}

This paper discusses CE specifically in connection to algorithmic recourse (AR). Commonplace definitions of AR include “an actionable set of changes a person can undertake in order to improve their outcome” \cite[p.1]{joshi2019revise}, “the ability of a person to obtain a desired outcome from a fixed model” \cite[p.10]{ustun2019actionable}, and “the systematic process of reversing unfavorable decisions by algorithms and bureaucracies across a range of counterfactual scenarios” \cite[p.284]{venkatasubramanian2020philosophical}.
Existing literature covers roughly two types of scenarios in which AR is typically sought. First, a decision subject may seek recourse for decisions that are correct yet unfavorable (e.g., a job applicant is rejected but wants to reapply in the future). Here, recourse might entail a specification of “actions that the agent would need to undertake to receive a more favorable decision from the algorithm” in the future \cite[p.288]{venkatasubramanian2020philosophical}.
Second, a decision subject may seek recourse for decisions that are unfavorable but also incorrect (e.g., the job applicant is rejected on the basis of erroneous or incomplete data). Here, recourse could entail reversing the decision and, depending on the case, possibly retraining the algorithm or changing or supplementing its input data \cite{cerrato2023correctability}. 

Recourse is generally understood to facilitate values like \textit{personal autonomy} and \textit{social trust}: when people (know that they) can reverse unfavorable decisions if needed, they can exercise their agency and plan for the future \cite{venkatasubramanian2020philosophical}. Our paper adopts the notion of AR as a \textit{modally robust good} advanced by \citet{venkatasubramanian2020philosophical} who appeal to \citeauthor{pettit2015robust}'s view according to which a good is modally robust if it remains valuable across possible worlds~\citep{pettit2015robust}. Recourse is valuable in a wide range range of possible situations and counterfactual circumstances, but arguably especially so in nonideal circumstances where decisions can be erroneous or normatively unfounded (e.g., unfair or arbitrary). Intuitively, the existence of effective mechanisms for appealing, contesting, and reversing decisions is particularly crucial for decision subjects where decision-makers may err or enact morally problematic decision policies.

CE is widely regarded as a promising means of providing recourse to decision subjects \cite{wachter2017counterfactual,verma2024counterfactual}. As explanations, CEs help decision subjects understand the reasons behind the decisions they receive, enabling them to also appeal and contest those decisions where needed. They can also highlight how the subject could receive a favorable decision in the future. Here, an actionable and minimal CE can be thought of as a recommendation that highlights a feasible set of actions the subject should undertake to receive a different (favorable) decision from the model at a later time. Relatedly, CE is seen as a promising approach to implementing the contentious right to explanation (e.g., \citet{wachter2017counterfactual}): ideally, they will express in an understandable format what features about the individual should have changed for the decision to change: ``had you done A instead of B, you would have received a favorable outcome". The idea that CEs should facilitate recourse \textit{reliably} in different circumstances appears to also motivate recent research that has developed methods to ensure that CEs remain \textit{robust} to changes in the data \cite{dominguezolmedo2022adversarial} and the model \cite{upadhyay2021robust}. Ensuring the technical robustness of CEs remains crucial in light of our findings. However, our results also demonstrate the need to attend to broader circumstantial variation---particularly, variation in upstream design choices and ML pipelines---to ensure that recourse mechanisms reliably facilitate their guiding values.

\subsection{Counterfactual Explanations for Algorithmic Recourse: Challenges and Limitations}

Our experiments are motivated by critical research examining the assumptions, challenges, and limitations of existing approaches to CE and AR. One key challenge for CE concerns the so-called ``Rashomon effect"~\citep{breiman2001statistical}: typically multiple non-identical counterfactuals can be computed even for a fixed model and output, and these may differ with respect to their validity and minimality. \citet{slack2021counterfactual} find that popular CE methods are also vulnerable to manipulation, lacking robustness against minor perturbations in the model. \citet{brughmans2024nice} test 12 different methods for CE and find high levels of disagreement between their respective explanations, suggesting that a particular counterfactual does not capture all relevant information and instead represents a partial and “unique point of view on the involved prediction logic” (p. 455). 
The idea that CE provides a restricted “view from somewhere” aligns with critical studies emphasizing that CEs are colored by ontological, epistemic, semantic, normative and empirical assumptions. For example,
\citet{barocas2020hidden} and \citet{kasirzadeh2021use} observe how CEs can be premised on contestable and empirically questionable assumptions pertaining to the relationship between model features and real-world actions, the commensurability and manipulability of features highlighted in counterfactuals, and the similarity between actual and counterfactual scenarios. Our empirical findings reinforce these critiques, demonstrating that CEs are sensitive to upstream assumptions and choices conditioning the generated counterfactuals.

Previously identified limitations of existing AR approaches represent form a second motivation for our study. Existing work highlights that standard AR frameworks adopt an \textit{individualistic} and\textit{ static} perspective to recourse \cite{venkatasubramanian2020philosophical,upadhyay2021robust,detoni2025time}. On the one hand, prevailing approaches tend to place responsibility for contesting and reversing decisions on the decision subject. This idea that ``recourse \textit{for} a person is necessarily recourse \textit{by} them" overlooks that recourse for the affected individual can require involving caretakers or legal representatives \cite[p.288]{venkatasubramanian2020philosophical}. On the other hand, standard AR approaches tend to  neglect temporal dynamics and instabilities, missing how updates to datasets and models, and changes in the underlying causal structure, can undermine the actionability and validity of recourse recommendations \cite{upadhyay2021robust,detoni2025time}. Actions that suffice for obtaining favorable outcomes at time \textit{t} may fail to achieve the same effect at \textit{t}+1, indicating that current approaches to AR fail to reliably facilitate decision reversal across changing circumstances.

The challenges and limitations discussed above intersect most clearly where CEs are used as a basis for recourse. For example, \citet[p.10]{slack2021counterfactual} note that the manipulability of CEs via minor data perturbations “bring[s] into the question the trustworthiness of counterfactual explanations as a tool to recommend recourse to algorithm stakeholders”. Similar concerns arise from changes due to distribution shifts and model updates as non-robust CEs may provide subjects seeking recourse with false hope of reversing their decision. These issues also problematize approaches to AR which view the decision subject as the locus of responsibility and the decision-maker as a ``benevolent paternalist" recommending what the subject should do. Indeed, while research has developed promising methods to generate \emph{robust} counterfactuals \cite{slack2021counterfactual,detoni2025time,upadhyay2021robust}, recourse recommendations that are invariant to changes in data and models do not avoid the problems stemming from this framing. Especially given the contentious nature of assumptions preceding the generation of CEs, their potential misalignment with real-world facts, and potential disagreement between multiple CEs for the same output, it appears further critical scrutiny of ``upstream" factors and choices is warranted. To this end, we analyze how upstream interventions at various stages of the ML pipeline affect CEs. 

\section{Example: Resume Screening}

We begin by introducing a running application example used throughout the remainder of this paper. We investigate candidate and resumé screening as a step of the hiring process which companies have an interest in automating \cite{sinha2020resume,lo2025ai} and which has been previously investigated critically in the literature due to its reliance on unvalidated psychometric measurements \cite{rhea2022external}. Commercial applications for screening we have reviewed appear to focus on computation of “fitness scores” of candidates to specific job positions, with a heavy emphasis on NLP and LLMs to generate both scores and ostensive explanations for these \cite{prescreenai2025}. Some companies even propose automated interviews according to scripts that can be given by domain experts and HR professionals \cite{alex2025}. In recent academic work, \citet{lo2025ai} propose the extraction and computation of five separate scores via LLM-based information extraction from resumes, dealing with i) self-evaluation of the candidate ii) job-relevant skills iii) previous work experience, iv) basic information and v) educational background, which can then be combined and employed to sort candidates. 

We design a hypothetical task of resume’ screening with the objective of adhering as much as possible to the characteristics of the tools and methodologies currently investigated and developed in this application context. However, we step back from LLM and RAG-based methodologies (Lo et al. 2025) as far as they are employed for candidate scoring directly. Our main interest is the production of CEs for screening decisions, and the production of counterfactuals in LLMs is not particularly mature and appears to suffer from issues of counterfactual validity \cite{youssef2024llms} in the sense discussed earlier in this paper. Instead, we operationalize the problem of resumé screening as one of supervised score-based ranking \cite{zehlike2022fairness}. We assume that a dataset $\mathcal{D} = \{(x_i, y_i)\}_{i=1}^n \subset \mathcal{X}^m \times \mathcal{Y}$ is given. Here, $\mathcal{X}^m$ represents a series of $m$ characteristics of individuals extracted from CVs such as previous job experience and education, whereas $\mathcal{Y}$ is the label space. We assume that an underlying, continuous fit construct $y^* \in [0, 1]$ exists, representing the probability score that a recruiter would, in principle, assess a candidate as the best one, and that the recorded labels $y_i$ are produced from it by an explicit business rule, so that the practical screening outcome is binary. 

Then, we define $F: \mathcal{X} \to \mathcal{Y}$ as a decision model trained on dataset $\mathcal{D}$ so to predict the potential fit of new unseen applicants which could be interviewed or not depending on the prediction of $F$. We then assume that individuals that are screened do not have access to $F$ but rather to a counterfactual explanation $\hat{x} = \mathcal{C}(F, x)$. Compared to usual formalizations of ranking, we do not explicitly represent a query $q$ as a feature vector and assume instead that the firm responsible to produce the model $F$ has a way to produce datasets $\mathcal{D}^q$ relevant to the specific query at hand. To clarify, imagine that the company implementing the resume’ screening service offers it to screen candidates for various positions, responsibilities, and competences.
The modeling approach usually adopted in the Information Retrieval literature here would be to represent a specific position to be filled as a query $q$ and assume that the model $F$ may take a vector representation of $q$ as an input. Here, we instead assume that depending on the position of interest, a relevant supervised dataset $\mathcal{D}$ is constructed which is then employed to fit the model $F$.
Thus, the good being distributed here by model $F$ is admittance to further interviewing, whether AI-based \cite{alex2025} or not, which we deem here a binary outcome and a potential target for a “why” question from candidates that have not been selected.
Thus, $F$ may also be analyzed as a binary classifier wherever relevant or useful.

\begin{figure*}
    \centering
    \includegraphics[width=0.5\linewidth]{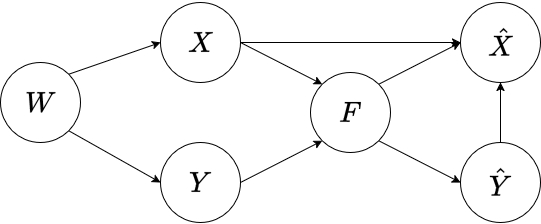}
    \caption{The structural causal model representing the data generating process. We assume that the world W is not observable, but that features $X$ and labels $Y$ may be obtained via measurement models. The trained model F similarly depends on certain design choices and produces labels $\hat{Y}$. The counterfactual examples $\hat{X}$ will ultimately depend on the interventions made by the developers regarding the SCM pictured here. We report structural equations and other variables in Equation \ref{eq:scm}.}
    \label{fig:scm}
\end{figure*}

We give the causal graph representing the assumed data generating process in Figure~\ref{fig:scm}. The features $X$ and labels $Y$ in $\mathcal{D}$ depend on measurement models $\phi_X$ and $\phi_Y$ which are non-observable by the candidates and only implicitly defined by the model-producing firm.

Before introducing the structural equations we fix notation. We use uppercase $X, Y$ for the random variables corresponding to the SCM nodes, and lowercase $x_i, y_i$ for their realizations in the dataset; their value spaces are $\mathcal{X}^m$ and $\mathcal{Y}$ respectively, so $x_i \in \mathcal{X}^m$ and $y_i \in \mathcal{Y}$. The dataset $\mathcal{D} = \{(x_i, y_i)\}_{i=1}^n$ is then an i.i.d.\ sample from the joint distribution that the SCM induces on $(X, Y)$. The recorded label is the post-business-rule label $Y$, not the latent fit construct $Y^*$: only the projection $Y = f_Y(Y^*, \phi_Y)$ is ever observable, which is precisely why $\phi_Y$ is one of the design choices we will intervene on. The same logic applies to $X$: the recorded feature vector is the output of the measurement model $\phi_X$ applied to the unobservable world $W$, and $\phi_X$ is similarly an explicit design choice rather than a fact of the world. How exactly these measurement models are operationalized has a downstream causal effect on the trained model $F$, which we make explicit in the structural equations of the SCM in Figure~\ref{fig:scm}:

\begin{align}
\label{eq:scm}
W & := N_W \\
X & := f_X(W, \phi_X, N_X)  \\
Y^* & := f_{Y^*}(W, N_{Y^*}) \\
Y & := f_Y(Y^*, \phi_Y) \\
F & := f_F(X, Y, M, C, K, N_F) \\ 
\hat{Y} & := F(X) \\
\hat{X} & := f_{\hat{X}}(X, \hat{Y}, F, \tau, N_{\hat{X}})
\end{align}

We use $F$ throughout for the trained model itself, and reserve $f_F$ for the training procedure that maps $(X, Y, M, C, K, N_F)$ to $F$; similarly $f_X, f_{\hat{X}}, f_Y, f_{Y^*}$ denote the structural functions producing the corresponding nodes. To describe the process, we assume that the features $X$ and raw labels $Y^*$ are measured noisily from the state of the world $W$. This may be understood as the process of obtaining certain measurements of skill, experience and fit for the position; we will be more concrete about the data setup in the Experiments section. We note that here we assume that the raw labels may be post-processed into $Y$ due to business requirements such as salary caps. The model training stage is represented by $f_F$ and we assume it to be causally influenced by a metric of success $M$, a model selection strategy $C$ and $K$ the model family or families being considered. The trained model $F$ then induces ranking labels $\hat{Y}$ which are used to generate counterfactuals $\hat{X}$ via the generation function $f_{\hat{X}}$. While counterfactual generation strategies have a plethora of hyperparameters, we focus here on $\tau$, the probability threshold that should be crossed by $F$ for a counterfactual $\hat{x}$ to be considered. In our experiments, we consider both most-minimal and most-valid counterfactuals by varying the threshold $\tau$.

\section{Experiments}\label{sec:experiments}

\paragraph{Data Sources.} The screening application described in the prior section motivates our choice of variables, but, as previously discussed, our experiments are not based on textual resumes. 
We use the American Community Survey (ACS) microdata sample obtained via the \texttt{folktables} library, treating each respondent as a candidate.
The ACS gives us actual population-level data for $X$, that is, the demographic, occupational and educational features that a screening firm would, on real resumes, have to reconstruct via parsing or LLM-based extraction. The latent fit construct $Y^*$ is however not so easily accessed: no public dataset that we are aware of connects features of this kind to recruiter judgments, and we have no easy way to bridge that gap. We therefore approximate $Y^*$ by an observed market signal: earnings, on the reading that wages reflect, imperfectly, the labour market's valuation of the same features that a screening model would weigh. In terms of the features $X$, we take a total of seven columns, with two continuous features and five discrete ones. The features contain diverse information, ranging from the type of college degree obtained (if any) to the number of working hours per week. We report the complete variable definitions in Table~\ref{app:tab:features} in the Appendix, due to space constraints. 

\paragraph{Experimental Design.} Our experimental setup is centered around the comparison of CE sets generated as a result of
an intervention on the causal graph of Figure~\ref{fig:scm} which we take to represent a ML pipeline. Thus, we are
interested in observing whether upstream interventions on the SCM, for instance intervening on the feature definitions and
labeling mechanisms, create significant variation on the generated CE sets. To understand whether a variation is  significant, we take a comparison with a downstream intervention on the same graph, that is, a change of the validity
threshold $\tau$. At a basic level, this value can be set at $\tau = 0.51$ to ensure that the trained model $F$ is confident
``just enough'' that a certain change in the feature values of an individual, resulting in the explanation or recommendation
$\hat{X}$, will change its output $\hat{Y}$. We then vary $\tau$ to higher values, and treat this downstream intervention as
our baseline: if upstream design decisions produce comparable or larger effects than directly tuning the explanation method,
this suggests that practitioners should attend to these decisions more carefully, and that such decisions should also be
considered as possible answers to the ``Why was this user not selected for screening?'' question. Specifically, we analyze the effect of the following interventions:
  \begin{itemize}
      \item Intervention 1: Label definition ($\text{do}(\phi_Y = s)$ for $s \in \{40\text{k}, 50\text{k}, 60\text{k},
  70\text{k}, 80\text{k}\}$). We vary the salary cap parameter that determines which individuals are labeled as positive.
  Specifically, we define $Y = 1$ if an individual is both in the top 10\% of earners and earns below the salary cap. This
  simulates a business constraint where the organization cannot afford candidates above a certain salary threshold.
      \item Intervention 2: Feature measurement ($\text{do}(\phi_X = p)$ for $p \in \{\text{for-profit}, \text{all-private},
  \text{private-self}, \text{non-gov}\}$). We vary the definition of ``private sector worker'' by including different
  combinations of employment classes: (i) private for-profit only, (ii) all private employers (for-profit and non-profit),
  (iii) private plus self-employed, and (iv) all non-government. This reflects ambiguity in how categorical concepts are
  operationalized from raw survey data.
      \item Intervention 3: Success metric ($\text{do}(M = m)$ for $m \in \{\text{accuracy}, \text{ROC-AUC}, \text{NDCG@100},
  \text{precision}, \text{recall}, \text{F1}\}$). We vary the metric used for model selection during cross-validation. This
  reflects choices made by ML practitioners when optimizing model performance.
      \item Intervention 4: Validation strategy ($\text{do}(C = c)$ for $c \in \{\text{holdout}, \text{5-fold},
  \text{10-fold}\}$). We vary the selection strategy for validation of the model $F$ from simple holdout (a 15\% test split
  followed by a 20\% validation split of the remainder, i.e.\ approximately 68/17/15 train/validation/test) to 5-fold and 10-fold cross-validation. Both this intervention and the preceding one tune the hyperparameters of a random forest classifier over a small grid that we report in the Appendix.
      \item Intervention 5: Stopping threshold ($\text{do}(\tau = t)$ for $t \in \{0.51, 0.55, 0.60, 0.65, 0.70\}$). We vary the confidence threshold used by the counterfactual generation algorithm\footnote{This is available as a parameter in DICE-kdtree, but not in NICE. For NICE, we simply reject CEs that are below the confidence.}. This serves as our baseline for comparison with the upstream interventions.
  \end{itemize}
We operationalize the ``impact of an intervention'' by measuring the sensitivity of counterfactuals across the separate causal graphs resulting from the intervention at hand:

\textbf{Definition (Explanation Sensitivity).} Let $\mathcal{S}$ be a structural causal model with design parameters $\xi = (\phi_X, \phi_Y, M, C, K, \tau)$. For a query instance $x$ with prediction $\hat{y} = F(x)$, let $\hat{X}_\xi = \mathcal{C}(F, x, \tau)$ denote the \emph{set} of counterfactual explanations generated under parameters $\xi$. The explanation sensitivity with respect to an intervention $\text{do}(\xi_i = \xi_i')$ on a single parameter $\xi_i$, holding all other parameters fixed, is:
\begin{equation}
S(\xi_i, \xi_i') = \mathbb{E}_{x}\left[ d(\hat{X}_{\xi}, \hat{X}_{\xi'}) \right]
\end{equation}
where $\xi'$ differs from $\xi$ only in component $i$, $d$ is a distance metric between CE sets on the feature space, and the expectation is taken over query instances $x$ for which the model gave a negative prediction $\hat{y} = 0$. 
In our experiments, we compute explanation sensitivity over sets of CEs generated by \textbf{DiCE-kdtree} \citep{mothilal2020explaining} and \textbf{NICE} \citep{brughmans2024nice}. Both methods are \emph{retrieval-based}: a counterfactual is selected from the training set rather than constructed by gradient-based optimization of a counterfactual loss.
We exclude optimization-based methods deliberately, as we observed in pilot experiments that their seed-to-seed variability swamped the effect of both upstream and downstream interventions.

We define the set distance $d(\hat{X}_\xi, \hat{X}_{\xi'})$ as a symmetric mean-of-nearest-neighbors distance, on which we further perform some rescaling to account for the different ranges of the features we employ. Continuous features are rescaled by their training-set median absolute deviation and categorical features contribute a 0/1 mismatch indicator. Full definitions are given in the Appendix.
Lastly, we note that the ``default setting'' of the SCM is $\xi_0 = (\phi_X^{\text{priv}}, \phi_Y^{60k}, M_{\text{acc}}, C_{5},
K_{\text{RF}}, \tau_{0.51})$: ``private sector'' restricted to private
for-profit employment, a salary cap of $\$60$k, accuracy as the model
selection metric, $5$-fold cross-validated grid search, a random-forest
hypothesis class,  $\tau=0.51$.

\paragraph{Research Questions.}

Our research questions are the following: 

\begin{itemize}
  \item \textbf{RQ1 (Sensitivity).} Are counterfactual explanations more
    \emph{sensitive} to upstream interventions on the SCM than to the
    downstream $\tau$ intervention?
  \item \textbf{RQ2 (Consistency).} Do upstream interventions produce more
    \emph{disagreement} between explanations than $\tau$ does?
\end{itemize}

\subsection{RQ1: Mechanism Sensitivity to the Four Interventions}\label{sec:sensitivity}

\begin{figure*}[ht]
    \centering
    \includegraphics[width=0.8\linewidth]{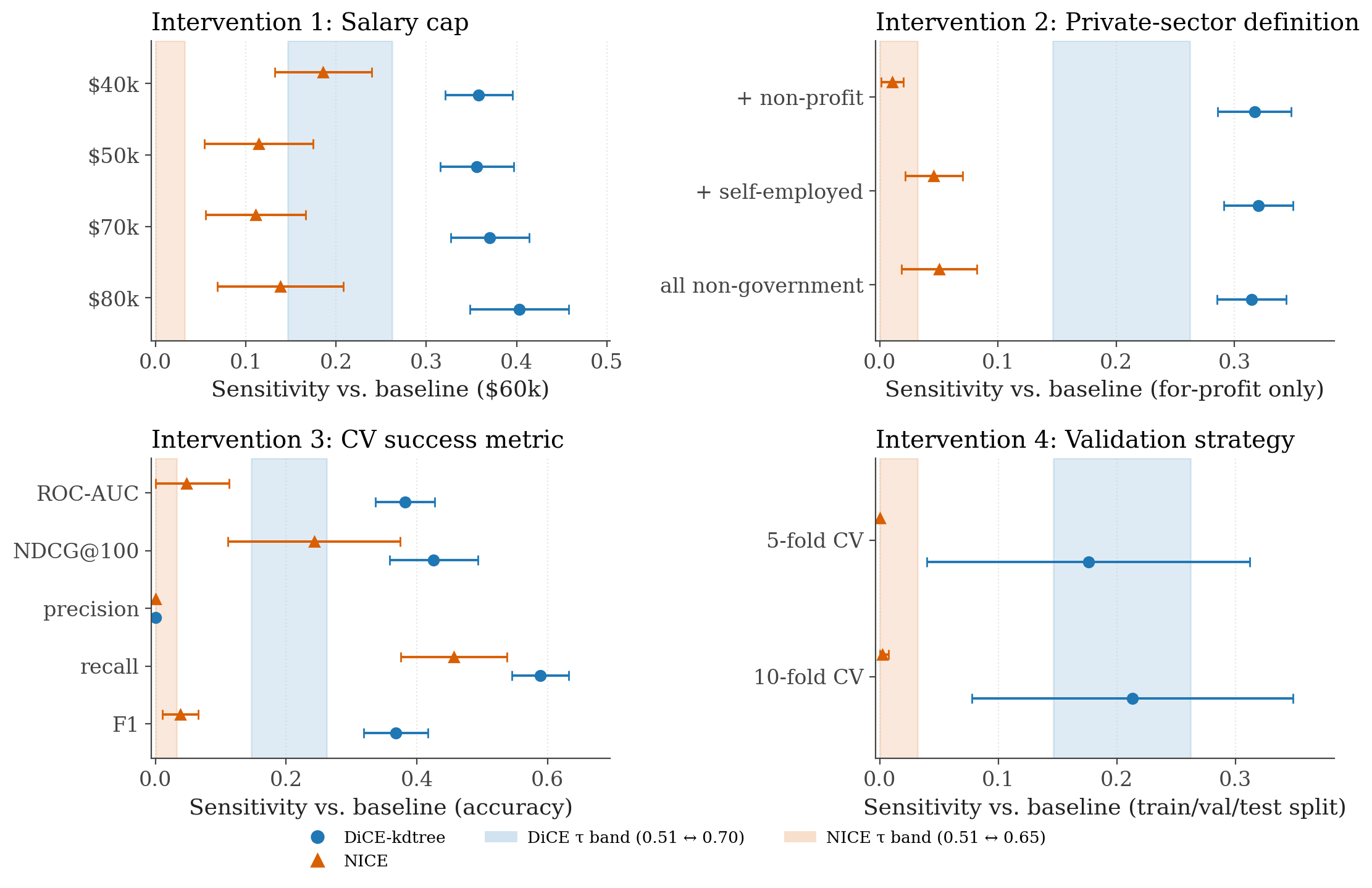}

    \caption{Mechanism sensitivity of counterfactual explanations to each of the five interventions on the SCM of Figure~\ref{fig:scm}. We show the effefct of intervening on the CE generating method ($\tau$) as a baseline (colored shaded band). Markers \emph{outside} the band indicate that the upstream intervention perturbs the explanation more than the explanation procedure's own validity threshold.}
    \label{fig:sensitivity-grid}
\end{figure*}

We report the five sensitivity interventions together in Figure~\ref{fig:sensitivity-grid}. Each panel shows, for one intervention, the mean symmetric set-distance between the strict counterfactual sets generated under the baseline level and each alternative level of that intervention, separately for DiCE-kdtree (circles) and NICE (triangles), with 95\% Student-$t$ CIs over ten seeds. The shaded band on each panel is the sensitivity of the generated CEs when intervening on $\tau$, where we take the maximum sensitivity we observed over $\tau$ interventions. A method whose marker sits outside this band has been moved more by the upstream intervention than it would have been by changing the validity threshold; a marker inside the band has been moved by an amount the threshold could already account for.

Three of the four interventions move both methods clearly outside the $\tau$ band. In the \textbf{salary-cap experiment} (Figure~\ref{fig:sensitivity-grid}, top left) sensitivity grows monotonically with the distance from the \$40k baseline, with caps at \$70k and \$80k producing distances well above the band for both methods.  We take this to indicate that business requirements such as salary cap are also viable answers to a ``why'' question. The same pattern holds for the operationalization of \textbf{private-sector worker} (Figure~\ref{fig:sensitivity-grid}, top right): all three alternatives to the for-profit-only definition DICE-kdtree outside the band, while NICE displays comparable or higher sensitivity across three settings, but within-band. 
The \textbf{success-metric experiment} (Figure~\ref{fig:sensitivity-grid}, bottom-left) is the most extreme of the four: switching from accuracy-based model selection to recall, ROC-AUC, NDCG@100, or F1 produces sensitivities that exceed the $\tau$. The only metric whose explanations remain inside the band is precision, which on this pipeline selects classifiers that overlap heavily with the accuracy-optimal model.
The \textbf{cross-validation experiment} (Figure~\ref{fig:sensitivity-grid}, bottom right) is the only case in which the upstream intervention does not dominate the $\tau$ hyperparameter: moving from a held-out train/val/test split to 5- or 10-fold cross-validation produces sensitivities that overlap with the $\tau$ band for both methods. This upstream intervention has therefore a comparable effect on the generated CEs as the downstream intervention on the generating method itself.
Taken together, Figure~\ref{fig:sensitivity-grid} answers \textbf{RQ1}. The choice of label definition ($\phi_Y$), feature measurement ($\phi_X$), and model-selection criterion ($M$) typically perturbs the explanation more than the explanation procedure's own validity threshold, while the choice of cross-validation strategy ($C$) is roughly comparable. This phenomenon holds across both retrieval-based CE methods.

\subsection{RQ2: Consistency of Explanations}\label{sec:consistency}

\begin{figure}
    \centering
    \includegraphics[width=\linewidth]{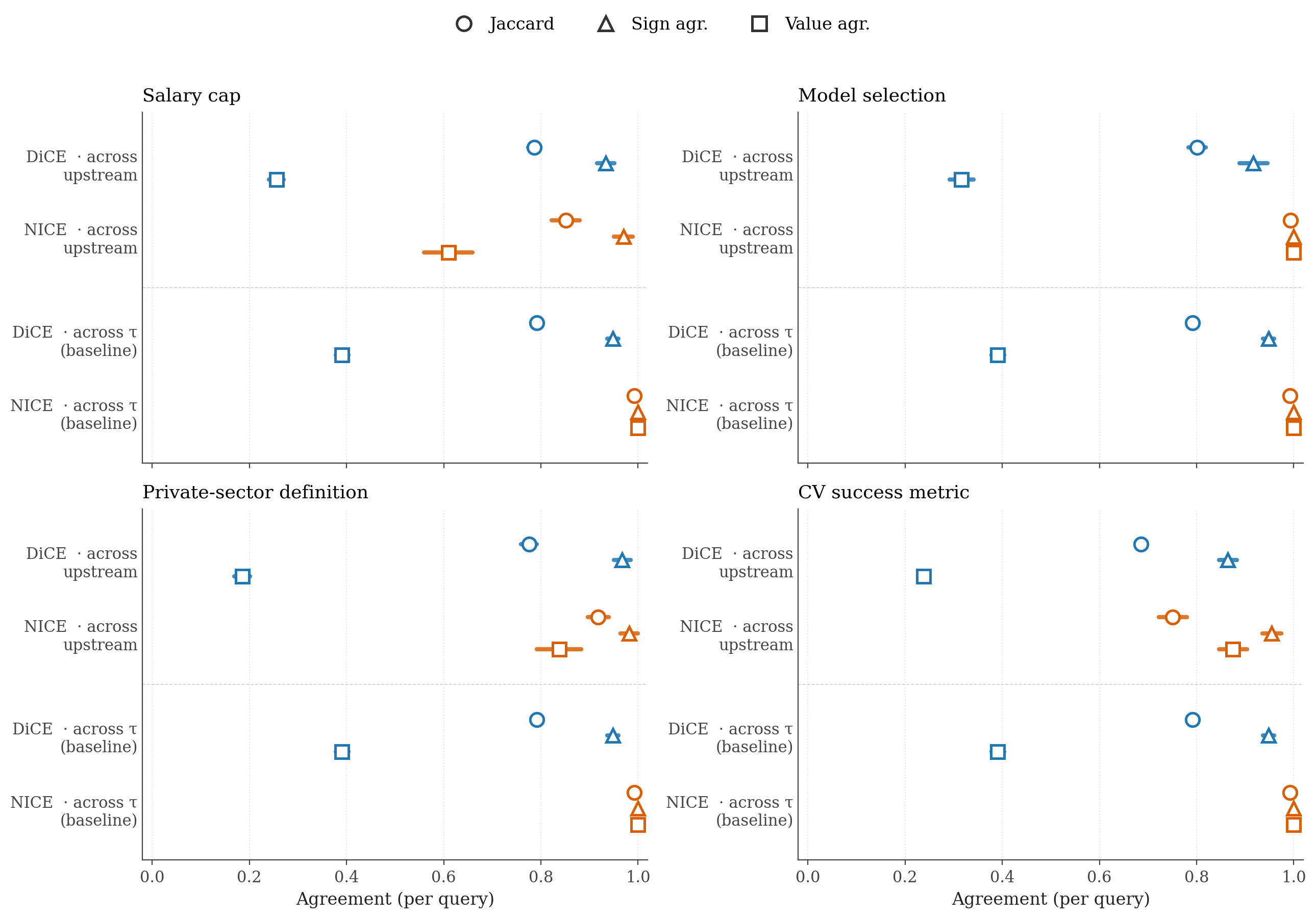}
    \caption{Within-method agreement under upstream interventions
    (top two rows of each panel) vs.\ the downstream $\tau$ baseline (bottom
    two rows). DiCE-kdtree in blue, NICE in orange. Markers are means with
    95\% Student-$t$ CIs over per-query scores; metrics are Jaccard ($\circ$),
    Sign Agreement on continuous features ($\triangle$), Value Agreement on
    categorical features ($\square$). Higher values indicate more agreement;
    an upstream marker to the \emph{left} of its $\tau$ counterpart means
    the upstream intervention produced more disagreement than $\tau$ does.}
    \label{fig:headline-consistency}
\end{figure}

The sensitivity analysis quantifies how far the CE set moves on average; a
complementary question, addressing \textbf{RQ2}, is whether upstream
interventions also produce more counterfactual \emph{disagreement} than
$\tau$ does. We follow \cite{brughmans2024disagreement} on the set of
features each CE changes relative to the query and report three per-query
agreement scores: their \textbf{Jaccard} agreement (do the same features change?), plus two extensions:
\textbf{Sign Agreement} on shared continuous features (same direction of
change?) and \textbf{Value Agreement} on shared categorical features (same
new value?). We report formal definitions and aggregation across multi-CE sets in the Appendix.
For each experiment, seed, and query instance $x$ with $\hat{y}=0$, we compute the three scores between the CE sets $\hat{X}_\xi(x)$ and
$\hat{X}_{\xi'}(x)$ for every pair of intervention levels $(\xi, \xi')$, holding all other parameters at the default setting $\xi_0$. 
We then average per-query scores and report them in Figure~\ref{fig:headline-consistency} as means with 95\% Student-$t$ confidence intervals over the ten seeds.

For three of four experiments, we observe DiCE-kdtree's mean Value Agreement drops from $0.39$ across different values of $\tau$ to between $0.18$ and $0.26$ under upstream interventions; NICE shows the same direction (from $1.00$ to between $0.61$ and $0.87$). The Jaccard agreement is also slightly lower for upstream interventions, while Sign Agreement is roughly equal both in downstream and upstream interventions.
The CV-folds panel (model selection, top right) is the exception, with upstream and $\tau$ at comparable agreement, which is consistent with the sensitivity result presented in the previous section. Our conclusion with regard to \textbf{RQ2}, is that upstream interventions may also impact the consistency of the explanation at the same level or slightly higher than intervening on $\tau$.

\section{Discussion}





Summarizing our results, we find that upstream interventions have a significant effect on the resulting counterfactuals.
Our operationalization of the sensitivity of CEs to interventions reveals that this effect is larger when compared to changing the validity threshold $\tau$ of the method, something that we observe in two separate retrieval-based CE generators (\textbf{RQ1}).
Compared to recent work from \citet{brughmans2024disagreement}, we show that disagreement in counterfactual explanations may also stem from upstream interventions (\textbf{RQ2}). 

We note here that we do not report this result intending to demonstrate a limitation of CEs \emph{per se}: compared to previous work (such as e.g. \citet{slack2021counterfactual}) we do not necessarily criticize the (lack of) robustness and manipulability of CEs to the upstream interventions we designed.
In fact, there exist methods to ameliorate the robustness of CEs to certain variations in the data and model (see e.g. \citet{virgolin2022on,jiang2024robust}), and we have no reason to believe that they could not be extended to account for some of all the upstream interventions we analyze in this paper.
Rather, our purpose with the quantitative evidence presented in the previous section is to substantiate our theoretical argument that several assumptions need to hold for CEs to represent an adequate ``why'', especially in applications such as justification and recourse.
The central assumption that we have operationalized with our experiments is the fixed model assumption, that is, that there is a specific model that best fits the data. 
While this may be the case, this is conditional on several actions from the decision-maker which are then \emph{not} provided as the underlying reason for a certain decision having being reached.
However, we have shown that certain decision-maker choices such as the model success metric, the labeling mechanism and the measurement model for certain features have significant impact on the generated CEs.

We now turn to the implications of our findings for real-life applications of CE, where they direct attention to the reasonableness of the fixed model assumption. Is the model stable over time? Is it legitimate in the sense that it is based on justifiable assumptions and design choices? The fixed model assumption can be assumed to be reasonable in some cases, where CE is used to explain outputs or debug models. Our findings are less directly relevant for such cases, accordingly, as the examined upstream interventions alter the model and its outputs. However, in many real-world settings data-generating processes evolve and models are updated, meaning pipeline stability cannot be assumed. Our findings have implications for such cases. Debugging signals produced with CE may fluctuate under upstream perturbations, for example, highlighting the need to distinguish issues arising from stable properties of the model from those stemming from data processing, labeling, and other contingent upstream factors. The observed sensitivity of CEs also underscores that CEs depend on upstream assumptions and choices that often remain open to normative and empirical contestation \cite{barocas2020hidden,kasirzadeh2021use}. Intuitively, \textit{why} a subject has received a particular decision depends not only on whether the subject meets certain criteria, but also on the deciding party's choices that shape the criteria---for example, feature measurement, labeling, success metrics, and more. Our results thus reinforce the idea that CEs are contestable artifacts of ML pipeline configurations which may themselves require critical scrutiny, an idea we examine next in connection to the provision of justification and recourse. In brief, we argue that CEs cannot reliably facilitate the guiding goals and values of these practices unless the legitimacy of upstream assumptions and design choices has been established.

\subsection{Implications for Justification}

The dependence of CEs on contestable upstream design and measurement choices has significant implications for their use in \textit{justifying} models and outputs in decision-making and auditing contexts. In practice, where models---and, by extension, assumptions and design choices shaping them---remain (or should remain) open to evaluation and revision, this dependence CEs marks a fundamental limit on their evidential and justificatory role. Because upstream choices also affect what counterfactuals are generated, CEs do not necessarily provide sufficient evidence for normative justification. Treating them as such inflates their evidential and justificatory role, and conflates conditional model behavior with adequate normative grounds for decisions. Meanwhile, it obfuscates agents' responsibilities by precluding consideration of the possibility that the model provider or deployer should justify and potentially revise their underlying assumptions and upstream choices.

Connecting to concerns of the manipulability of CEs (Slack et al. 2021), we note a related risk: CEs may be misinterpreted as model-\textit{in}dependent evidence or used selectively to shirk responsibility and public accountability. To illustrate, consider a stylized example in which our hypothetical recruiter is suspected of indirect age discrimination, and the burden of proof has shifted onto them. To rebut the suspicion, the recruiter conducts  a counterfactual fairness analysis by perturbing applicants' age to measure differences in outcomes. Suppose the analysis shows quantitatively non-negligible age-based disparities. However, the recruiter argues that they reflect age-correlated differences in qualifications and skills, and that they are also proportional to those differences. This exemplifies the risk that CEs may be misinterpreted as evidence of causality \cite{karimi2021algorithmic}. But the recruiter's CEs \textit{presuppose} a causal structure over the data as opposed to providing evidence for a specific causal explanation of the disparities (e.g., that age causes differences in the screened features). The generated CEs cannot establish that the disparities are justified, accordingly, as supplementary evidence is required. 

The CEs may provide evidence of the disparities' proportionality to base-rate differences, supporting the recruiter's claim. Nonetheless, this remains \textit{insufficient} for justification in the case at hand. The recruiter must also demonstrate that they have legitimate aim, and that the preferred model is necessary and appropriate for achieving that aim. Importantly, however, this cannot be demonstrated under the assumption of a fixed model. Rather, comparison to reasonable alternatives is required, including but not limited to\textit{ }\textit{alternative models} produced under different specifications of the SCM. This point is echoed in recent works stating that decision-makers have a legal duty to search for and implement alternative models which exhibit comparable performance with less disparity \cite{black2024less,laufer2025constitutes}. If CEs are to facilitate the required comparisons, they must notably be produced for \textit{multiple} models trained on the same data. Our results show that, besides model selection, choices of labels and metrics can also affect CEs. Thus, they further underline the need to scrutinize upstream assumptions and choices, and to assess the comparability, commensurability, and reliability of the compared CEs \cite{barocas2020hidden,kasirzadeh2021use}.  

\subsection{Implications for Recourse}

Our findings have significant implications for using CEs as a basis for recourse, lending support to critiques described in prior sections. On the one hand, they raise concerns about the (long-term) robustness and actionability of CEs \textit{qua} recourse recommendations beyond concerns of data and model drift \cite{upadhyay2021robust,detoni2025time}. How decisions could be reversed depends not only on subjects' features and the chosen decision threshold, but also on upstream elements of the ML pipeline. Accordingly, CEs produced under different assumptions and design choices are also likely to differ in properties such as minimality and actionability. Given the substantial changes observed under alternative feature measurements and model selection metrics, for instance, one can expect notable differences in the associated \textit{costs} of achieving a favorable outcome. Where the stability of the ML pipeline is not guaranteed, technical robustness under a fixed model fails to ensure that CEs facilitate the values that recourse should facilitate in a modally robust manner. Rather, if CEs are to reliably recommend feasible pathways to favorable outcomes under circumstantial variation, their robustness to upstream perturbations must be ensured. 

Our results call into questions whether recourse recommendations could also remain \textit{normatively underdetermined} by standard evaluation criteria. If CEs for models produced under distinct but equally defensible upstream choices and assumptions describe incompatible recommendations for recourse (e.g., changing different features), then there may be no uniquely identifiable action the decision subject should have taken to achieve a favorable outcome. This possibility underscores our previous point about the order of justification: upstream choices must be relatively settled, normatively speaking, prior to evaluating CEs for recourse. This also gives reason to expand the normative basis of evaluating recourse recommendations with additional desiderata such as fairness or diversity \cite{boxer2025realizing,laugel2023achieving}. 

On the other hand, our findings also challenge the prevailing individualistic and unilateral framing of AR \cite{venkatasubramanian2020philosophical}. Because CEs bracket upstream assumptions, choices, and their consequences, they represent partial and \textit{incomplete} answers to why a subject received the decision but also to how that decision could be reversed. This abstraction from the actions and responsibilities of model developers and/or deployers implicitly forecloses consideration of the possibility that \textit{they} ought to have acted otherwise, and that the normatively preferable path to favorable outcomes may be to revise the model or decision policy itself. In the language of modal robustness, CE as a recourse mechanism does not reliably facilitate trust or subjects' autonomy across counterfactual cases where better design and measurement choices could have been made ( either in the empirical or normative sense). Thus, where the model's stability or normative justification remains unsettled, CEs should not be taken at face value. Treating them as individualized advice for reversing decisions obscures their status as artifacts of organizations' design and governance choices, and risks encouraging affected subjects to comply with potentially problematic policies.

Taken together, these points highlight the need to move beyond the simplistic conception of AR as a unilateral process in which decision-makers benevolently recommend actions to affected decision-subjects. They serve as an important reminder that AR is an inherently \textit{relational practice} involving both decision subjects and decision-makers as parties whose behavior may require revision. In practice, evaluating whether recourse mechanisms reliably facilitates goods like personal autonomy and peace of mind requires considering \textit{all} parties to the institutional relation along with their respective possibilities and obligations for action. While design choices impact the CEs decision-subjects receive, it may be impossible for subjects to reverse their decisions \textit{through} contesting such choices absent appropriate transparency. Beyond explainability, effective and robust recourse thus requires broader measures of ``algorithmic transparency"---including but not limited to data-sharing structures and documentation trails \cite{gebru_datasheets_2021}---that enable critical scrutiny and informed appeal and contestation. The relational perspective also avoids insulating the deciding party's design choices from critical scrutiny without undermining individual responsibility. While it recognizes that decision-subjects can exercise agency to alter their situation, it also does not unfairly place onto the subjects the burden of simply adapting to institutional discretion.

\section{Conclusions}

In this paper, we provided an empirical and theoretical analysis of the limits of CEs. Complementing existing work on their robustness, we measured the sensitivity of CEs to systematic ``upstream interventions” across the ML pipeline and showed that CEs generated with the same method vary substantially in response to design choices made prior to the point of explanation. Our experiments illustrate that CEs track many factors in the ML pipeline which condition the model to which a CE method is applied. While our analysis focuses on a single CE method and metric (an extended version of Jaccard agreement), we expect similar patterns to hold more broadly. We encourage future research to test this with other local explanation methods (including alternative CE methods), alternative robustness measures, and recourse-relevant metrics such as diversity and actionability.

Building on our findings and prior critical research on CE and AR, we argued that CE can play only a limited role in real-life practices of explanation, debugging, justification, and recourse provision. In particular, we argued that the potential temporal instability of key design choices, along with the contestability of their underlying assumptions, should inform the evaluation and practical application of CE methods. As model-dependent artifacts, CEs offer, at best, partial justification for models or decisions, and are prone to misinterpretation and strategic misuse. Our findings also challenge individualistic and unilateral understandings of AR. Recourse is relational and involves multiple parties, and thus effective recourse mechanisms must address the possibility that organizations' design choices may be empirically or normatively untenable and in need of critical scrutiny and potential revision. Therefore, CEs alone, absent broader ``algorithmic transparency", remain insufficient to secure core values like public accountability, trust, and agency. 

\section*{Acknowledgements}
The authors gratefully acknowledge the support of the ``TOPML: Trading Off Non-Functional Properties of Machine Learning'' project funded by the Carl Zeiss Foundation, grant number P2021-02-014.

\bibliography{references}

\clearpage
\begin{table*}[h]
\caption{The ACS Microdata Sample columns we employed in our experiments. More detailed definitions are available at the official ACS documentation: \url{https://www2.census.gov/programs-surveys/acs/tech_docs/pums/data_dict/PUMS_Data_Dictionary_2018.pdf}.\label{app:tab:features}
}
\centering
\begin{tabular}{@{}lll@{}}
\toprule
Column Name & Definition                                                          & Type        \\ \midrule
JWTR        & Means of transportation to work                                     & Categorical \\
WKW         & Weeks worked during past 12 months                                  & Continuous  \\
ENG         & Ability to Speak English                                            & Categorical \\
SCIENGP     & Achieved a Degree in Science and Engineering, by the NSF Definition & Categorical \\
COW         & Class of worker                                                     & Categorical \\
SOC         & Standard Occupational Classification code                           & Categorical \\
JWMNP       & Travel time to work                                                 & Continuous  \\ \bottomrule
\end{tabular}
\end{table*}

\section{Hyperparameter Selection}

We report the hyperparameter grids and defaults we employ in Interventions 3 and 4 (model selection strategy, model success metric) in Table~\ref{app:tab:rf-grid}. The ``Default'' column shows the non-interventional choice, that is, the SCM setting $\xi_0$. 

\begin{table}[h]
  \centering
  \caption{Random-forest hyperparameter grid searched during model selection. Cross-validation uses stratified $C$-fold splits
  with the success metric $M$; the validation strategy $C$ and metric $M$ are themselves interventions in
  Section~\ref{sec:experiments}. All other RF parameters are left at their \texttt{scikit-learn}
  defaults.\label{app:tab:rf-grid}}
  \begin{tabular}{@{}lll@{}}
  \toprule
  Hyperparameter & Grid values &  Default \\ \midrule
  \texttt{n\_estimators}       & $\{50,\ 100,\ 200\}$       & $100$ \\
  \texttt{max\_depth}          & $\{5,\ 10,\ \text{None}\}$ & None \\
  \texttt{min\_samples\_split} & $\{2,\ 5\}$                & $2$ \\ \bottomrule
  \end{tabular}
\end{table}

\section{Counterfactual distance metrics}\label{app:distance}

We measure the sensitivity of CE sets to an intervention by averaging a set-level distance $d(\hat{X}_\xi, \hat{X}_{\xi'})$ over query instances. The set distance is built in two layers: a per-CE cost $c(x, x')$ on the feature space, and a symmetric aggregator over sets of CEs.

\paragraph{Per-CE cost.} For two feature vectors $x, x'$ over continuous features $\mathcal{F}_{\text{cont}}$ and categorical
 features $\mathcal{F}_{\text{cat}}$, we use a mixed-type cost
\begin{equation}
c(x, x') = \sum_{j \in \mathcal{F}_{\text{cont}}} \frac{\lvert x_j - x'_j \rvert}{\operatorname{MAD}_j(X_{\text{train}})} +
\sum_{j \in \mathcal{F}_{\text{cat}}} \mathbf{1}\!\left[x_j \neq x'_j\right],
\end{equation}
where $\operatorname{MAD}_j$ is the median absolute deviation of feature $j$ estimated on the run's training set. 
The rescaling makes continuous contributions unit-free and outlier-robust; categorical features contribute a 0/1 mismatch indicator. The MAD is fitted per seed on that seed's training set, so $c$ is run-specific.

\paragraph{Set-level distance.} For two CE sets $A, B$ associated with the same query, we use the symmetric
mean-of-nearest-neighbours distance
\begin{equation}
d(A, B) = \tfrac{1}{2}\!\left( \frac{1}{|A|}\sum_{a \in A} \min_{b \in B} c(a, b) + \frac{1}{|B|}\sum_{b \in B} \min_{a \in
A} c(a, b) \right).
\end{equation}
This is a softened, averaging variant of the Hausdorff distance: replacing each mean by a max recovers Hausdorff. We use the mean form because retrieval-based methods can return CE sets of mildly different sizes ($K_{\text{DiCE}} = 5$, $K_{\text{NICE}} = 3$), and a worst-case aggregator would tend to over-weight the single most discordant CE in either set. If either set is empty, $d$ is undefined and the corresponding query is dropped from the across-query average.

\paragraph{Aggregation across queries and seeds.} For each intervention pair $(\xi, \xi')$, method, and seed, we compute
$d(\hat{X}_\xi(x), \hat{X}_{\xi'}(x))$ at every query $x$ with $\hat{y} = 0$ and take the mean over queries. Reported sensitivities are means across seeds with 95\% Student-$t$ confidence intervals.

\section{Consistency metrics: full definitions}\label{app:consistency}

This appendix provides the formal definitions, equation numbers, and design choices underlying the consistency analysis proposed in our experiments.

\subsection{Explanation set}

We propose a slightly different notation for queries and feature vectors here compared to our experimental section, for convenience and clarity. Let $x$ be a query instance with feature vector $x \in \mathbb{R}^{|\mathcal{F}|}$ over the feature schema $\mathcal{F}$ of dataset $D$, and let $\hat{x}$ be a counterfactual produced for $x$. We define the \emph{explanation set} of $\hat{x}$ relative to $x$ is

\begin{equation}
E(\hat{x}, x) = \{ j \in \mathcal{F} : \hat{x}_j \neq x_j \},
\end{equation}

i.e.\ the subset of features that the counterfactual changes relative to the query. Continuous features are compared with an absolute tolerance of $10^{-9}$; categorical features are compared as strings.

\subsection{Pairwise metrics}

For two counterfactuals $\hat{x}_a, \hat{x}_b$ of the same query $x$, with explanation sets $E_a = E(\hat{x}_a, x)$ and $E_b = E(\hat{x}_b, x)$, the paper defines several disagreement metrics. We use the symmetric \textbf{Jaccard agreement}:

\begin{align}
\text{Jaccard}(E_a, E_b) &= \frac{|E_a \cap E_b|}{|E_a \cup E_b|} \quad
\end{align}

Jaccard is $1$ when the two CEs change exactly the same set of features and $0$ when they are disjoint.

The formulation from \cite{brughmans2024disagreement} treats two CEs as "agreeing on feature $j$" whenever both change $j$, regardless of how. Their analysis focuses on understanding the possibility that a malicious actor would include an arbitrary feature into an explanation as a means of assuaging recourse. As a consequence, a CE that changes (for instance) \texttt{occupation} from \emph{cook} to \emph{chef} agrees with one that changes it from \emph{cook} to \emph{lawyer} only at the level of "occupation was changed". 
Here, we are instead interested in understanding the overall direction of change suggested by two CEs across different interventions, and whether the direction is in some sense similar. To recover the directional information for continuous features and the exact-value information for categorical features we define two complementary scalars, both intersected with $E_a \cap E_b$ so that they only score features both CEs changed.

Let $\mathcal{F}_{\text{cont}}, \mathcal{F}_{\text{cat}}$ partition $\mathcal{F}$ into continuous and categorical features.

\paragraph{Sign Agreement (continuous).} We compute the sign agreement of continuous variables in a CE as follows:

\begin{align*}
\mathrm{SA}(\hat{x}_a, \hat{x}_b, x) &= \frac{1}{|M|} \sum_{j \in M} \mathbf{1}\!\left[ \mathrm{sign}(\hat{x}_{a,j} - x_j) = \mathrm{sign}(\hat{x}_{b,j} - x_j) \right]. \\
J &= E_a \cap E_b \cap \mathcal{F}_{\text{cont}}
\end{align*}

$\mathrm{SA} = 1$ when the two CEs always push shared continuous features in the same direction.

\paragraph{Value Agreement (categorical).} Closes the "agree-on-feature-but-not-value" gap for categorical features:

\begin{align}
\mathrm{VA}(\hat{x}_a, \hat{x}_b, x) &= \frac{1}{|M|} \sum_{j \in M} \mathbf{1}\!\left[ \hat{x}_{a,j} = \hat{x}_{b,j} \right]. \\
M &= E_a \cap E_b \cap \mathcal{F}_{\text{cat}}
\end{align}

$\mathrm{VA} = 1$ when the two CEs always pick the same new value for shared categorical features.

Both metrics are undefined when the corresponding intersection is empty (no continuous or no categorical feature changed by both CEs). In that case, the original query vector is dropped from the overall computation of the metric.

\subsection{Aggregation over multi-CE sets}\label{app:cf-set-aggregation}

In our setup each (run, method, threshold, intervention, query) tuple yields a CE \emph{set} of $k \geq 1$ counterfactuals (with $k = 4$ for DiCE-kdtree and $k = 2$ for NICE in our configuration). 
We note that reducing each set to a single representative CE (e.g. with the median for continuous features, mode for categoricals) and applying the single-CE metrics, would smooth out differences in the CEs.
Instead, for two CE sets $A = \{\hat{x}^A_1, \dots, \hat{x}^A_{k_A}\}$ and $B = \{\hat{x}^B_1, \dots, \hat{x}^B_{k_B}\}$ generated for the same query $x$ under different conditions, we define the set-level metric as the average over the Cartesian product:

\begin{equation}
\mathrm{Jaccard}(A, B, x) = \frac{1}{k_A k_B} \sum_{i=1}^{k_A} \sum_{l=1}^{k_B} \mathrm{Jaccard}\!\left(E(\hat{x}^A_i, x), E(\hat{x}^B_l, x)\right),
\end{equation}

and analogously for FD, SA, and VA.

\end{document}